\documentclass[12pt,amsfonts]{article}
\begin{document}
\def\p {{\partial}}
\def\n {{\nu}}
\def\m {{\mu}}
\def\a {{\alpha}}
\def\bt {{\beta}}
\def\f {{\phi}}
\def\th {{\theta}}
\def\g {{\gamma}}
\def\eps {{\epsilon}}
\def\e {{\psi}}
\def\la {{\lambda}}
\def\na {{\nabla}}
\def\k {\chi}
\def\bn {\begin{eqnarray}}
\def\en {\end{eqnarray}}
\title{Hamilton-Jacobi formulation of systems within Caputo's fractional derivative
} \maketitle
\begin{center}
\vspace{0.2cm} \textbf{Eqab M.Rabei}\footnote{E-mail:eqabrabei@yahoo.com}\\
 Department of
Physics,Mutah University, Karak, Jordan\\
\vspace{0.2cm} \textbf{Ibtesam Almayteh}\\
 Department of
Physics, Mutah University, Karak, Jordan\\

\vspace{0.2cm} \textbf{Sami I. Muslih}\footnote{E-mail:
smuslih@ictp.trieste.it}\\Department of Physics, Al-Azhar
University, Gaza, Palestine\\

\author\textbf{Dumitru Baleanu}\footnote{On leave of absence from
Institute of Space Sciences, P.O.BOX, MG-23, R 76900,
Magurele-Bucharest, Romania,E-mails: dumitru@cankaya.edu.tr,
baleanu@venus.nipne.ro}\\
Department of Mathematics and Computer Sciences, Faculty of Arts
and Sciences,$\c{C}ankaya$ University- 06530, Ankara, Turkey \\

\end{center}

PACS:

\begin{abstract}
In this paper we develop a fractional Hamilton-Jacobi formulation
for discrete systems in terms of fractional Caputo derivatives.
The fractional action function is obtained and the solutions of
the equations of motion are recovered.  An example is studied in
details.
\end{abstract}

\section{Introduction}

Fractional calculus  generalized the classical calculus and it has
many important applications in various fields of science and
engineering [1-10]. These applications include classical and
quantum mechanics, field theory, and optimal control [11-30]
formulated mostly in terms of Riemann-Liouville (RL) and Caputo
fractional derivatives. In contrast with RL derivative,Caputo
derivative of a constant is zero, and for a fractional
differential equation defined in terms of Caputo derivatives the
standard boundary conditions are well defined.
 Therefore, this kind of fractional derivative gained importance  among
engineers and scientists.

 Recently, the fractional Hamiltonian
formulations were investigated for fractional discrete and
continuous systems  in terms of RL and Caputo derivatives [22-37].
Even more recently, a fractional Hamilton-Jacobi formalism within
RL derivative was proposed in [38].

As it well known  the classical Hamilton–Jacobi equation
represents a reformulation of classical mechanics which equivalent
to other formulations such as Newton's laws of motion, Lagrangian
mechanics and Hamiltonian mechanics. In addition, the
Hamilton–Jacobi equation is  useful in finding the conserved
quantities for mechanical systems, which may be possible even when
the mechanical problem itself cannot be solved completely. The
Hamilton-Jacobi theory represents the only formulation of
mechanics in which the motion of a particle can be represented as
a wave.

On the other hand in the area of fractional mechanics the
investigation of the fractional Hamilton-Jacobi equation is still
at the beginning of its development.

Due to the above mentioned reasons the formulation of a fractional
Hamilton-Jacobi  within Caputo fractional derivatives is an
interesting issue to be investigated.

The plan of this paper is as follows:

In section two the basic mathematical tools are briefly described.
In section three the fractional Hamilton equations within Caputo's
derivative are mentioned. Section four contains the fractional
Hamilton-Jacobi formulation by using Caputo's derivative. An
illustrative example is analyzed in section five. Finally, section
six contains our conclusions.

\section{Basic definitions}
In the following we briefly present some fundamental definitions
used in the previous section.

 The left and the right Riemann-Liouville
and Caputo fractional derivatives are defined as follows:\\

\noindent {\em The Left Riemann-Liouville Fractional Derivative}
\begin{equation}
_aD_t^\alpha f(t) = \frac{1}{\Gamma (n-\alpha)} \left(
\frac{d}{dt} \right)^n \int_a^t (t-\tau)^{n-\alpha-1} f(\tau)
d\tau,
\end{equation} 

\noindent {\em The Right Riemann-Liouville Fractional Derivative}
\begin{equation}
_tD_b^\alpha f(t) = \frac{1}{\Gamma (n-\alpha)}
\left(-\frac{d}{dt} \right)^n \int_t^b (\tau-t)^{n-\alpha-1}
f(\tau) d\tau,
\end{equation} 

The corresponding Caputo's fractional derivatives are defined as
follows:\\

 \noindent {\em The Left Caputo Fractional Derivative}
\begin{equation}
_a^CD_t^\alpha f(t) = \frac{1}{\Gamma (n-\alpha)} \int_a^t
(t-\tau)^{n-\alpha-1} \left( \frac{d}{d\tau} \right)^n f(\tau)
d\tau ,
\end{equation} 

\noindent and\\

{\em The Right Caputo Fractional Derivative}
\begin{equation}
_t^CD_b^\alpha f(t) = \frac{1}{\Gamma (n-\alpha)} \int_t^b
(\tau-t)^{n-\alpha-1} \left(-\frac{d}{d\tau} \right)^n f(\tau)
d\tau ,\end{equation} 

\noindent where $\alpha$ represents the order of the derivative
such that $n-1 < \alpha < n$.

In  \cite{Agrawalcankaya1} the fractional Euler-Lagrange equations
are obtained. We present briefly the main result obtained in the
following (for more details see \cite{Agrawalcankaya1}).

 {\em Let $J[q]$ be a functional of the
form
\begin{equation}
J[q] = \int_a^b L(t, q, \,_a^CD_t^\alpha q, \,_t^CD_b^\beta q) dt
\end{equation}
\noindent {\em  where $0 < \alpha, \beta < 1$} and defined on the
set of functions $y(x)$ which have continuous LCFD of order
$\alpha$ and RCFD of order $\beta$ in $[a, b]$. Then a necessary
condition for $J[q]$ to have an extremum for a given function
$q(t)$ is that $q(t)$ satisfy the generalized Euler-Lagrange
equation given by }
\begin{equation}\label{eleq}
\frac{\partial L}{\partial q} + \,_tD_b^\alpha \frac{\partial L
}{\partial \,_a^CD_t^\alpha q } + \,_aD_t^\beta \frac{\partial L
}{\partial \,_t^CD_b^\beta q} = 0, \hspace{0.2in} t \in [a, \,\,
b]
\end{equation} 
\noindent {\em and the transversality conditions given by}
\begin{equation}
\left[ _tD_b^{\alpha-1} ( \frac{\partial L }{\partial
\,_a^CD_t^\alpha q } ) - _aD_t^{\beta-1} ( \frac{\partial L
}{\partial \,_t^CD_b^\beta q } ) \right] \eta(t) |_a^b = 0.
\end{equation}

\section{Fractional Hamiltonian formulation}
 In this section we briefly present the Hamiltonian formulation within
Caputo's fractional derivatives.

 Let us consider the fractional Lagrangian as given below
\begin{equation}\label{unu}
L(q,_a^CD_t^\alpha q,_t^C D_b^\beta q,t ), \hspace{0.2in}
0<\alpha,\beta< 1.
\end{equation}
By using (\ref{unu}) we define the canonical momenta $p_\alpha$
and $p_\beta$ as follows
\begin{equation}\label{doi}
p_\alpha=\frac{\partial L}{\partial _a^CD_t^\alpha q },
\hspace{0.2in} p_\beta=\frac{\partial L}{\partial _t^CD_b^\beta q
}.
\end{equation}
Making use of (\ref{unu}) and (\ref{doi}) we define the fractional
canonical Hamiltonian as
\begin{equation}\label{trei}
H=p_\alpha \, {_a^CD_t^\alpha q}+{p_{\beta}} \, _t^CD_b^\beta q-
L.
\end{equation}

The fractional Hamilton equations are obtained as follows [37]

\begin{eqnarray}\label{ul}
\frac{\partial H}{\partial t}=-\frac{\partial L}{\partial t},
\hspace{0.1in} \frac{\partial H}{\partial p_{\alpha}}= \,
_a^CD_t^\alpha q, \hspace{0.1in} \frac{\partial H}{\partial
p_{\beta}}= \, _t^CD_b^\beta q, \hspace{0.1in} \frac{\partial
H}{\partial q}= \, _tD_b^\alpha p_{\alpha} + \,_aD_t^\beta
p_{\beta}.
\end{eqnarray}

\section{Fractional Hamilton-Jacobi formulation}
In this section, we construct the Hamilton-Jacobi formulation
within Caputo fractional derivatives. According to [37], the
fractional Hamiltonian is defined in equation (10). Now, let
consider the canonical transformations with the generating
function $F_2( {_a^CD_t^{\alpha-1} q} ,  _t^CD_b^{\beta-1} q,
P_\a, P_\bt, t) = S$. The new Hamiltonian will take the form
\begin{equation}
K= P_\alpha \, {_a^CD_t^\alpha Q}+{P_{\beta}} \, _t^CD_b^\beta Q-
L(Q, {_a^CD_t^\alpha Q}, _t^CD_b^\beta Q).
\end{equation}
The variations of the following integrals
\begin{equation}
\delta \int_{t_1}^{t_2}(P_\alpha \, {_a^CD_t^\alpha Q}+{P_{\beta}}
\, _t^CD_b^\beta Q -  K)dt =0, ~~~\delta \int_{t_1}^{t_2}(p_\alpha
\, {_a^CD_t^\alpha q}+{p_{\beta}} \, _t^CD_b^\beta q- H)dt =0,
\end{equation}
are vanishing, then we obtain the relation between the
Hamiltonians $H$ and $K$ as follows
\begin{equation}
p_\alpha \, {_a^CD_t^\alpha q}+{p_{\beta}} \, _t^CD_b^\beta q- H =
P_\alpha \, {_a^CD_t^\alpha Q}+{P_{\beta}} \, _t^CD_b^\beta Q -  K
+ \frac{d F}{d t},
\end{equation}
where the function $F$ is given as
\begin{equation}
F= S({_a^CD_t^{\alpha-1} q} , _t^CD_b^{\beta-1} q, P_\a, P_\bt, t)
- P_{\a}{_a^CD_t^{\alpha-1} Q}- P_{\bt} {_t^CD_b^{\beta-1} Q}.
\end{equation}
Substituting this  function $F$ in  equation   (14), we obtain the
following equations \bn && {_a^CD_t^{\alpha-1} Q} = \frac{\p S}{\p
P_\a}, ~~~~~~~{_t^CD_b^{\bt-1} Q} = \frac{\p S}{\p P_\bt}\\
&& p_a = \frac{\p S}{\p {_a^CD_t^{\alpha-1} q}}, ~~~~~~~~~p_\bt =
\frac{\p S}{\p {_t^CD_b^{\bt-1} q}},\\
&& K = H +\frac{\p S}{\p t}. \en In the case that the new
variables $(Q, P_{\a}, P_{\bt})$ are constants in time, then
$K=0$. One can easily show that action function can be put in the
form.
\begin{equation}
S= \int_{t_1}^{t_2} L dt.
\end{equation}
For time-independent Hamiltonian $H$, the action function $S$ can
be put in the form
\begin{equation}
S=W_1(_a^CD_t^{\alpha-1}q, E_1)  + W_2(_t^CD_t^{\bt -1}q, E_2) +
f(E_1, E_2, t),
\end{equation}
here, $W$ is called the Hamilton's characteristic function and
$f(E_1, E_2, t) = -Et$, where $E_1= P_{\a}$, $E_2= P_{\bt}$  and
$E= E_1 + E_2$.

Making use of equation (22), and after some simple manipulations,
we obtain \bn && {_a^CD_t^{\alpha-1} Q} = \frac{\p S}{\p
P_\a}=\lambda_1, ~~~~~~~{_t^CD_b^{\bt-1} Q} = \frac{\p S}{\p P_\bt}=\lambda_2\\
&& p_\a = \frac{\p W_1}{\p {_a^CD_t^{\alpha-1} q}}, ~~~~~~~~~p_\bt
=
\frac{\p W_2}{\p {_t^CD_b^{\bt-1} q}},\\
&& \frac{\p S}{\p t} =- H =- E. \en and the Hamilton-Jacobi
partial differential equation for fractional system reads as
\begin{equation}
\frac{\p S}{\p t} + H = 0.
\end{equation}
Now the solutions of equation (20), give the values of $W_1$ and
$W_2$ as
\begin{equation}
W_1 = \int p_\alpha d {_a^CD_t^{\alpha-1} q}, ~~~~W_2 = \int p_\bt
d {_t^CD_b^{\bt -1} q}.
\end{equation}
\section{Illustrative example}
In this section, we shall present an example to demonstrate the
application of the formulation developed in the previous section.
 Let us consider the following Hamiltonian [36]
 \begin{equation}
 H = \frac{1}{2}{p_{\a}}^{2} + q.
 \end{equation}

The fractional Hamilton-Jacobi equation is given by
\begin{equation}
\frac{1}{2}{p_{\a}}^{2} + q - E_1 =0. ~~~~~p_\bt =0.
\end{equation}
Making use of equation (22), we get
\begin{equation}
\left[\frac{\p W_1}{\p {_a^CD_t^{\alpha-1} q}}\right]^{2} + q -
E_1 =0,
\end{equation}
which has the following solution
\begin{equation}
W_1=\sqrt{2(E_1 - q)} {_a^CD_t^{\alpha-1} q},
\end{equation}
this equation leads us to obtain $p_\a$ and $S$ respectively as
\bn&& p_{\a}= \sqrt{2(E_1 - q)}\\
&&S=\sqrt{2(E_1 - q)} {_a^CD_t^{\alpha-1} q}- E_1t.\en Again
making use of equations (21), (31), we arrive at
\begin{equation}
_a^CD_t^{\alpha-1} Q = \frac{\p S}{\p t} = \frac{1}{\sqrt{2(E_1 -
q)}}{_a^CD_t^{\alpha-1} q} -t =\lambda_1.
\end{equation}
Equation (32), can be easily solved to obtain
\begin{equation}
{_a^CD_t^{\alpha-1} q} = \sqrt{2(E_1 - q)}(t +\lambda_1),
\end{equation}
or
\begin{equation}
{_a^CD_t^{\alpha} q} = \sqrt{2(E_1 - q)}=p_{\a}.
\end{equation}
Taking the Riemman-Liouville derivative of equation (34), we
obtain
\begin{equation}
{_t D_b^{\bt}} _a^CD_t^{\alpha} q =1.
\end{equation}
This result is in exact agreement with that obtained if we use the
fractional Hamiltonian formulation.

\section{Conclusions}

In this paper, we have obtained the Hamilton-Jacobi partial
differential equation within Caputo's fractional derivative.
Finding the action function leads us to obtain the solutions of
equations of motion. An example was investigated and the solutions
of the equations of motion are in exact agreement with those
obtained by using the Hamiltonian formulation. The advantages of
using the method presented in this paper, is that we can easily
obtain the action function, which is the essential part to obtain
the path integral quantization for any mechanical fractional
system, and this topic is now under investigation.

\end{document}